Title: Small polaron confinement revisited
Author: Mladen Georgiev (Institute of Solid State Physics, Bulgarian Academy of Science,
   1784 Sofia, Bulgaria)
Comments: 25 pages and 14 figures, all pdf format.
Subj-class: physics

Confinement processes arranging small polarons into insulating periodic structures above certain conversion temperatures are considered. The binding energy of a structure is an extension of Van-der-Waals' (VdW) molecular force to colossal magnitudes inherent of VdW energies in dilute media. VdW binding reduces to usual magnitudes when the interlevel energy gap drops down as pressure is increased. VdW pairing of vibronic small polarons arises from phonon coupling to two-level orbital systems. Vibronic (Jahn-Teller) polarons associating inherent electric & magnetic dipoles coupled to external fields lead to colossal field effects. Confinements based on Poisson-Boltzmann equations of electrostatic strings and Poisson-Fermi ones are deduced.. Formation and demolition mechanisms of small-polaron structures in manganites and oxocuprates are proposed.

1. Introduction

Small vibronic polarons have been playng an increasingly essential role in interpreting solid state observations and also in substantiating solid state hypotheses. Namely, small polarons have been considered appropriate for explaining confinement and field effects observed in solid state materials such as colossal resistance manganites [1] and high-temperature superconducting oxocuprates [2]. In as much as the basic metal ions in both of them ($Mn^{3+}$ and $Cu^{2+}$, respectively) are Jahn-Teller ions, the small polarons have often been assumed to appear in their vibronic form: Jahn-Teller (JT) or pseudo-Jahn-Teller (pJT) [3]. It is to be stressed that one of the earliest expectations for a viable high-temperature superconducting mechanism has relied on the presumed role of JT and pJT ions [4].

Below, we consider these two illuminating examples of materials and species that possibly exhibit vibronic small polaron confinement at $T \geq T_C$: *viz.* manganites and oxocuprates. The Colossal Field Resistance (CFR) and High-Temperature Superconductivity (HTSC) are among the discoveries of modern solid state physics during the past quarter of a century.

CFR is observed as a multi-order-of-magnitude drop in field resistance as a CMR or CER active material is placed in an external magnetic (electric) field. The classical CMR (CER) material is manganese oxide (manganite) of the $La_{1-x}Sr_xMnO_3$ family, all of the perovskite $LaMnO_3$ structure, as shown in Figure 1(a) [1]. Figure 1(a) also depicts even parity vibrations of the oxygen octahedron around a $Mn^{3+}$ ion which can give rise to Jahn-Teller (JT) distortions ($E_g$ mode), leading possibly to JT-polarons. In the absence of an external field, the conductivity σ of a CMR material is metallic below a conversion temperature $T_C$ (Curie point), at which σ is vanishing, the material turning insulating at $T \geq T_C$. It is believed that the conduction currents are carried by large polarons at $T \leq T_C$ which turn small at $T \geq T_C$ that are

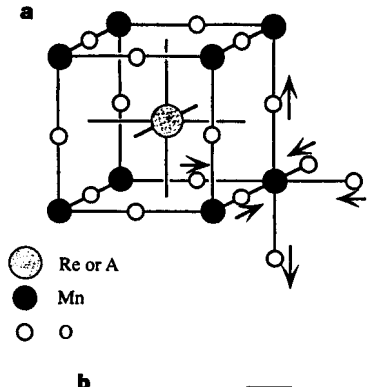

Figure 1(a): The Manganite unit cell. The big ball at the body center is a rare-earth ion, usually $La^{3+}$. The magnetic $Mn^{3+}$ ions are situated at the vertexes of the cube, the compensating $O^{2-}$ ions are edge centered. Each manganese ion is at the center of an $6O^{2-}$ oxygen octahedron whose ions vibrate in an even parity $E_g$ – like mode. The compound formula is $RE_{1-x}A_xMnO_4$ where RE is a rare earth, A is a divalent substitutional cation dopant introduced to implant holes.

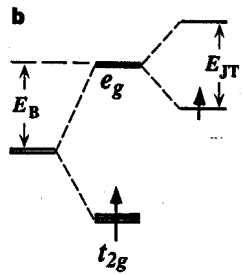

Figure 1(c): The manganese energy levels and their splitting. The original Mn d-electron quintet splits into a $t_{2g}$ triplet and a $e_g$ doublet in the crystalline field. The doublet, if singly occupied, is further split through coupling to the $E_g$ Jahn-Teller mode, the splitting amounts to the Jahn-Teller energy. Both diagrams are taken from Millis' paper in Reference [2].

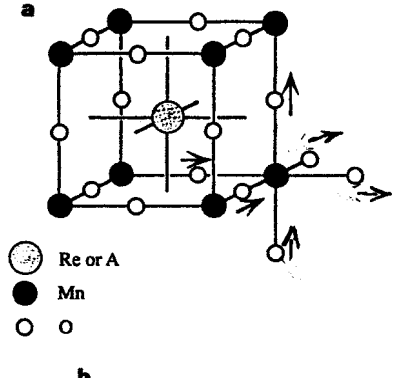

Figure 1(d): Showing the $T_{1u}$- like odd-parity vibration of the oxygen octahedron. Coupling of this mode to a pair of opposite-parity electronic states may result in an off-centering of the oxygen octahedron from the normal Mn site. This will give rise to the breaking down of the inversion site symmetry and the consequent occurrence of an electric dipole. In addition, the octahedron may reorientate through jumping over the off-center sites in an orbital rotation to give rise to orbital magnetic dipoles. The obtained picture apparently contains most of the features needed for the structure to couple to the external fields.

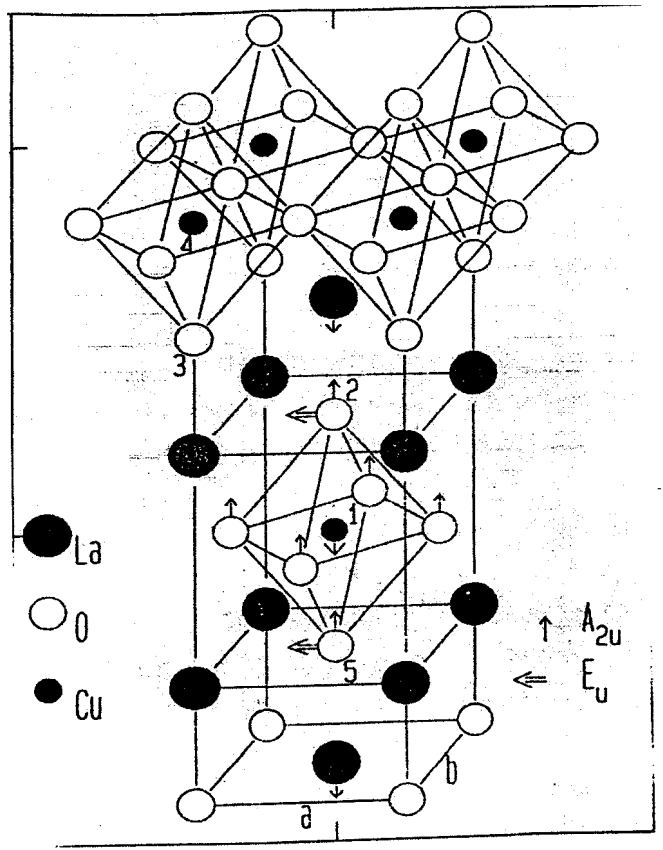

Figure 1(b): The HTSC La$_{2-x}$Sr$_x$CuO$_4$ unit cell. Note the occurrence of several types of symmetries or symmetry violations: JT (vertical) elongation along the c-axis and the in-plane orthorhombic distortion ($a \neq b$). These general features are common for the manganites and the oxocuprates. At the center is the basic unit of CuO$_6$ octahedron with its six oxygens (four planar and two apical ones). The La$_2$O$_3$ rocksalt layers hosting the apex oxygens can be viewed separately.(From A.G. Andreev's PhD Thesis.)

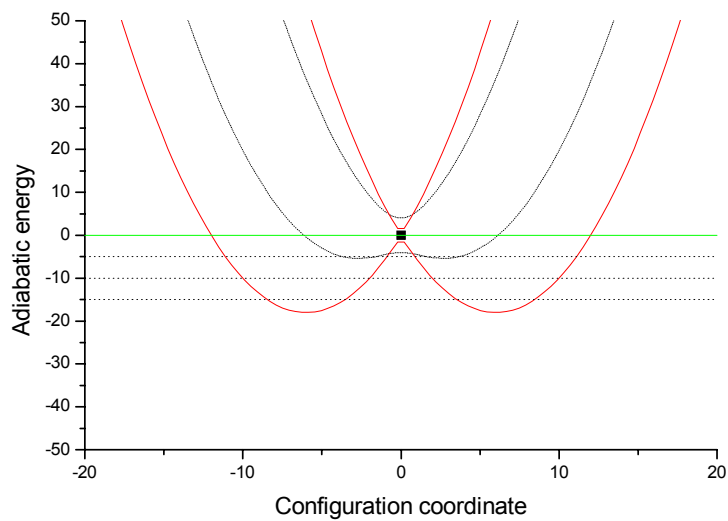

Figure 2: Pairs of adiabatic potentials for small polarons (solid lines) and large polarons (dotted lines). The small polarons are locally confined to very narrow squeezed bands (dotted) while the large polarons are only weakly bound moving in wide bands through the diagram center (solid reference).

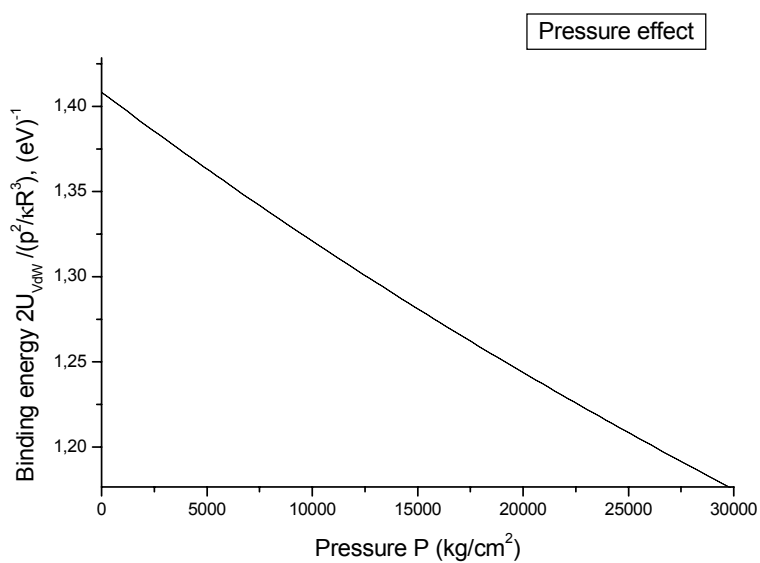

Figure 3: Calculated pressure effect on the colossal Van der Waals binding energy using data by Reference 13. The molecular system moves to the right as tha universe evolves in time.

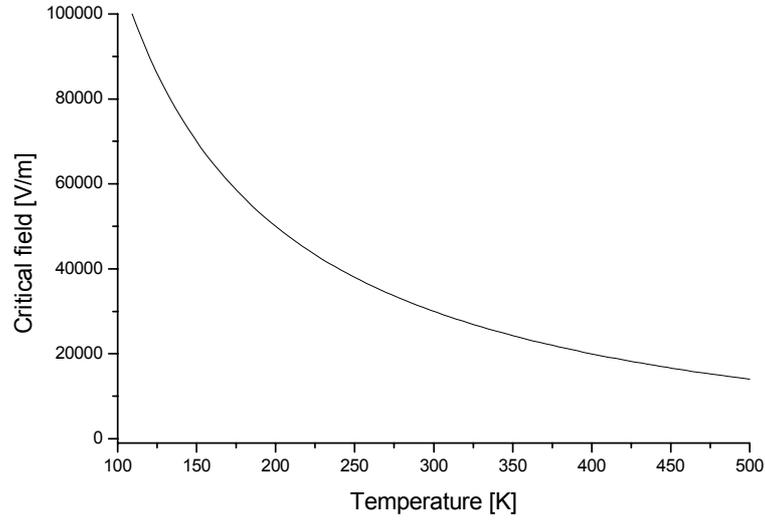

Figure 4: The temperature dependence of the critical electric field from $F_c = (p_{12}^3/\kappa^2) \times (\alpha_M/R^6)/18k_BT - F_\infty$, where the symbols are explained in Sections 2.2 and 3, while the intercept $F_\infty = \Delta E(0)/p_{12}$ is related to the zero-field tunneling splitting. From the available value of $F_\infty \sim 10^4$ V/m at $p_{12} \sim 1$ eÅ, we get $\Delta E(0) \sim 1$ μeV which is not at all an epicyclic estimate. Finally, we use the calculation of Section 3 to get $F_c = 1.20 \times 10^7 / T - 10^4$ [V/m] to calculate a tentative temperature dependence of the critical field.

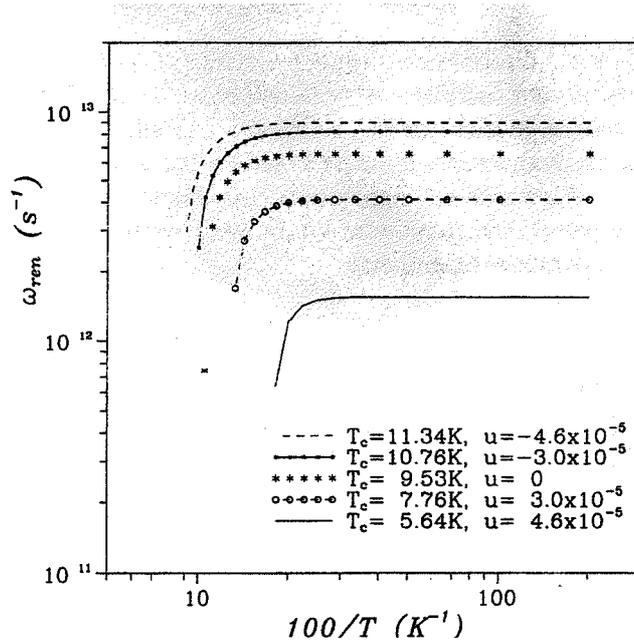

Figure 5: The temperature dependences of the renormalized vibrational frequencies at various pair interaction strengths u. $T_C$ are the corresponding low-symmetry to high-symmetry conversion temperatures from equation (59).

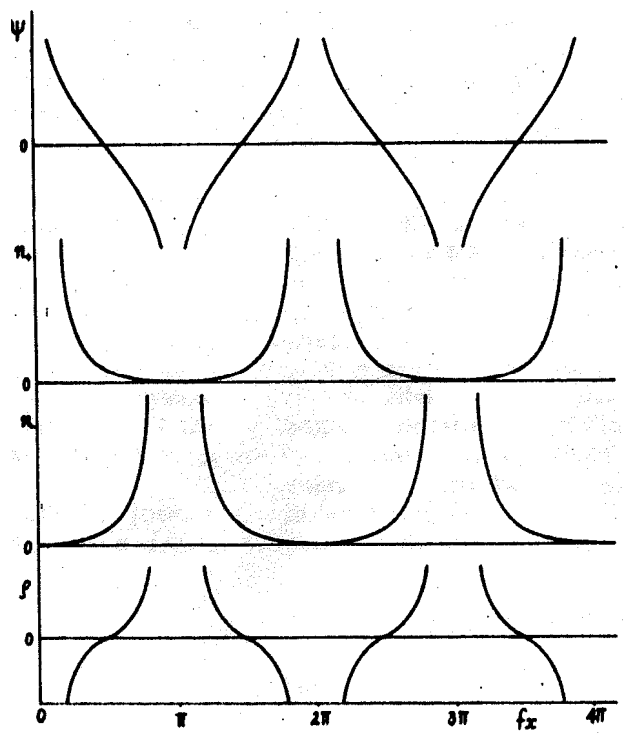

Figure 6: Basic quantities pertaining to 2D PB plates (sheets). From top to bottom: electrostatic potential $\psi$, negative particle density $n^-$, positive particle density $n^+$, space charge $\rho$. (From Reference 12.)

far less mobile [4]. The small polarons form kind of a superlattice which is melted away by the external field bringing back the material to the original conductive (large polaron) state. In effect, the applied external field displaces the conversion temperature $T_C$ towards the higher T along the temperature axis. The superlattice melting (order→disorder) has recently been proven by experiment, at least in its CER form [5].

There are a lot of classic HTSC material families which are to be described separately. In the single-layered family $La_{2-x}Sr_xCuO_4$, the basic structural element is the $CuO_6$ octahedron with two apex oxygens O(A), four planar oxygens O(P) and a copper Cu(P) at the center of the plate. Besides, neighboring octahedra are isolated from each other along the c-axis by $La_2O_3$ rocksalt layers. The interlayer coupling, however, essential for the superconductivity is due to the bridging function of the apex oxygens. Elevating the superconducting temperature constitutes the main goal of both theory and experiment. We refer the reader to a multitude of good books and reviews on the physics of HTSC materials. See Figure 1(b) for an illustration of the $La_{2-x}Sr_xCuO_4$ single-layered superconducting family.

The small polarons must each carry both electric and magnetic dipoles in order to couple to the external fields. In looking for a species, we remind of the off-center small polarons which are both electrostatically polarizable and carrying an orbital magnetic dipole [6]. We had earlier proposed that dispersive pairing forces could arise within a field of vibronic small polarons to produce a binding mechanism. Now we shall pursue the matter to some more detail as regards the colossal resistance effects in manganites.

The conversion mechanism is not yet clear though it suggests a field-dependent electron-phonon coupling constant G. In addition, G is strong though unconventional since one might normally expect small bound polarons to build up preceding the formation along the temperature axis of large polarons (else free carriers). One possible reason has been given in that the ferromagnetic (conductive) alignment of electron spins comes from an increased bandwidth which suppresses the small-polaron formation below the Curie temperature [1]. Nevertheless, the above mechanism may be expected to work better for Holstein's small polarons and less so for Jahn-Teller (JT)- or Pseudo-Jahn-Teller (pJT)- small polarons [4].

The free $Mn^{3+}$ ion 3d-electron quintet is split by the crystalline field to a higher-lying $e_g$ doublet and a lower-lying $t_{2g}$ triplet. If the $e_g$ doublet is singly occupied, it may couple to an even parity $E_g$ vibrational mode to produce a co-symmetric JT distortion of the surrounding $6O^{2-}$ oxygen octahedron, as in Figure 1(c). If not occupied,. then a doublet level may mix vibronically through, say, an $T_{1u}$ odd-mode coupling with some odd-parity lower- lying level to produce a pJT off-center distortion of the $6O^{2-}$ octahedron at the Mn site.

On the other hand, one could hardly devise a simple magnetic field dependent electron-phonon coupling, which is the realm of electrostatics. $Mn^{3+}$ is a magnetic ion but, in addition, the Mn site may carry an orbital magnetic dipole if its $6O^{2-}$ oxygen frame is slightly displaced off center in a way shown in Figure 1(d) ($T_{1u}$ mode). As the off-center Mn site rotates jumping around the normal lattice site, the off-site dipole varies in both magnitude and direction and its average vanishes. Still, the off-site Mn is polarizable electrostatically and may give rise to a Van der Waals (VdW) coupling between neighboring Mn off-sites. The latter dispersive-type coupling may result in the formation of a periodic ferromagnetic or antiferromagnetic off-site orbital structure to host the small polarons. Another important conjecture is that $Mn^{3+}$ is a Jahn-Teller ion (cf. Figure 1(c)) leading to a vertical elongation of the oxygen octahedron (cf. Figure 1(a)). However, this distortion does not contribute to the Mn site electric dipole and thereby to the small-polaron superlattice or to the field coupling strength.

The small-polaron lattice, whether off-site or otherwise, seems to play an essential role in the field-dependent drop of electrical resistance. Namely, the rise in polaron conductivity being heralded by the melting of the small-polaron lattice, the conductivity conversion at $T_C$ can also be interpreted as one from a solid polaron lattice to its liquid melt form (see Figure 2 for the vibronic potentials distinguishing small from large polarons). It is implied that the conductivity is hindered by the periodic small-polaron structure, which binds the carriers, and is stimulated by this structure getting soft. The solid (small polarons) to liquid (large polarons) conversion is the heart of the CMR (CER) mechanism. Lattice coupling have long been considered essential for the field-enhancement of electric conductivity of manganites [1].

Turning back to the small-polaron periodic structure, we stress that in our model it can be visualized as an array of consecutive left-hand and right-hand rotors kept intact by the dispersive VdW chemical bond. The dispersive force is enhanced by the lattice coupling reducing largely the electron gap energy, originally the order of an eV, to negligible values. This reduction called Holstein's band narrowing enhances the linear off-site polarizability and is the chief factor for the occurrence of a colossal VdW binding [6]. Now, if the sample with a small polaron superlattice, sustained by the colossal binding obliged to the enhanced linear polarizability, is placed in an electric field of sufficient strength, as it seems to be the case, the linear polarizability will be substituted for by the nonlinear or field-dependent polarizability which is considerably lower. In effect the polarizability lowering at the small-polaron sites would undoubtedly reduce the chemical binding therein causing the periodic structure to disintegrate.

The above being a plausible scenario for CER, the CMR mechanism requires additional consideration.. We only suggest that the orbital magnetic dipoles of superlattice polarons should likely be aligned in the antiferromagnetic manner, by alternating left-hand and right-hand rotations (chirality). As an external magnetic field would tend to turn the dipole arrangements ferromagnetic, this would in effect increase the energy of the structure turning it less stable. In this way an external magnetic field of sufficient strength may again bring about the disintegration of the superlattice.

The stability of a VdW structure off-field is the main sustaining factor for the present vibronic model. It is based on the observed pressure dependence of the interlevel energy gap as shown in Figure 3. On the other hand, Figure 4 shows the temperature dependence of the critical CER electric field as derived in the earlier work.

While looking for experimental confirmations at that time, news came in of an experimental report of a colossal electroresistance [5]. It confirmed the occurrence of a solid to liquid conversion when the field applied to manganites was sufficiently high.

---

2. Field coupling to a small-polaron superlattice

2.1. Field-off analysis

For a better account of the underlying physics, we repeat arguments of an earlier publication [7]. Consider a molecular cluster with two nearly-degenerate opposite-parity electronic eigenstates $\psi_1$ and $\psi_2$ of eigen-energies $\varepsilon_1$ and $\varepsilon_2$, respectively. If now $\mathbf{p}_{12} = <\psi_2|\ e\mathbf{r}\ |\psi_1>$ is the electric dipole mixing these states, the electrostatic polarizability of the two-level system is

$$\alpha_{el} = (1/3)\ p_{12}^2\ /\ |\varepsilon_2 - \varepsilon_1| \qquad (1)$$

The two-level electrostatic polarizability (1) brings about a VdW electronic pairing energy

$$U_{VdWel} = \tfrac{1}{2}\ \Delta E_{gap}\ (\alpha_{el}\ /\ \kappa\ R_{ij}^3)^2 = (p_{12}^4/18\kappa^2 E_{gap})(1/R_{ij}^6) \qquad (2)$$

$\kappa$ is the dielectric constant of the medium, $R_{ij}$ is the pair separation, $E_{gap} = |\varepsilon_2 - \varepsilon_1|$ is the interlevel energy gap. The pairing energy (2) arises from electrostatic interactions within the *bare two-level system*. However, eq. (2) is a 2$^{nd}$ order result and may not hold true otherwise.

We subsequently assume that the two electronic states are coupled to mix *vibronically* by an odd-parity intermolecular vibration Q through the *pseudo-Jahn-Teller effect*. As a result of the vibronic mixing the original interlevel spacing $E_{gap}$ reduces to

$$\Delta E_{gap} = E_{gap}\ \exp(-2E_{JT}/\hbar\omega) \qquad (3)$$

(Holstein effect) where $E_{JT} = G^2/2K$ is Jahn-Teller's energy, $\omega \equiv \omega_{ren} = \omega_{bare}\sqrt{[1 - (E_{gap}/4E_{JT})^2]}$ is the renormalized frequency of the coupled vibration, $\omega_{bare}$ being the bare vibrational frequency, $\hbar = h/2\pi$. The electrostatic polarizability of the *squeezed two-level system* turns in

$$\alpha_{vib}^0 = (1/3)\ p_{vib}^2\ /\ |\Delta E_{gap}|\ , \qquad (4)$$

termed *vibronic polarizability*, with

$$p_{vib} = p_{12}\ \sqrt{[1 - (E_{gap}/4E_{JT})^2]} \qquad (5)$$

standing for the *vibronic electric dipole*. The vibronic mixing effects give rise to a *vibronic VdW pairing energy* in lieu of equation (2) [4,6]:

$$U_{VdWvib} = \tfrac{1}{2}\ \Delta E_{gap}\ (\alpha_{vib}^0\ /\ \kappa\ R_{ij}^3)^2$$

$$= (p_{12}^4/18\kappa^2 \Delta E_{gap})(1/R_{ij}^6)[1 - (E_{gap}/4E_{JT})^2]^2 \qquad (6)$$

(See References [4,6,7] for more details.) The vibronic polarizability $\alpha_{vib}$ is temperature-dependent ($\alpha_{vib}^0$ is its low-temperature value).. For a molecular system [8]:

$$\alpha_{vib}(T) = \alpha_{vib}^0\ \tanh(|\Delta E\ |/k_B T) \qquad (7)$$

From (6) we obtain the attractive part of the electrostatic *VdW binding energy* $V_{vib}$ *of a vibronic two-level system* at 0 K:

$$V_{vib}^0 = \tfrac{1}{2}\ \Delta E_{gap}\ (\alpha_{vib}^0/\kappa)^2\ \Sigma_{ij}\ R_{ij}^{-6} \qquad (8)$$

The vibronic energy (8) is to be compared with the attractive part of the *VdW binding energy* $V_{el}$ *of an electronic system in the uncoupled two-level state*:

$$V_{el} = \tfrac{1}{2} E_{gap}(\alpha_{exc}/\kappa)^2 \sum R_{ij}^{-6} \qquad (9)$$

Taking the ratio of (8) to (9) we get

$$V_{vib}^0 / V_{el} = [1 - (E_{gap}/4E_{JT})^2]^2 \exp(2E_{JT}/\hbar\omega) \qquad (10)$$

It is the large exponential term (reciprocal Holstein reduction factor) that makes $V_{vib} / V_{el} \gg 1$. We also note that for small polarons $4E_{JT} \gg E_{gap}$ Equation (10) justifies terming the vibronic VdW interaction energy '*colossal*'.

Equations (9) and (8) for the binding energy of a two-level system may be translated into a binding energy of a small-polaron superlattice on introducing the respective pairing energies (2) and (6), provided each pair is counted once. Assuming additivity of the VdW interaction energy this gives

$$V_{vib}^0 = \tfrac{1}{2}\Delta E_{gap}(\alpha_{vib}^0/\kappa)^2 (\alpha_M/R^6) \qquad (11)$$

where $\alpha_M = \sum_{ij} (R/R_{ij})^6$ is the sum constant of the small-polaron superlattice. In as much as the Mn sublattice is simple cubic crystallographically, so is the small-polaron superlattice with a lattice constant R in equation (11). Performing the summation we get $\alpha_M = 6(1 + \tfrac{1}{4} + 1/8 + \ldots) \approx 8.25$.

2.2. Field-on analysis

We refer the reader to earlier publications for a survey of the coupling of lattice oscillators to an external field [6,7]. The relevant analysis leads to the following field coupling terms: $H'_E = -\mathbf{p}.\mathbf{E}$ and $H'_H = -\mathbf{\mu}.\mathbf{H}$ where **p** and **μ** are the electric and magnetic dipoles, respectively, pertaining to a single oscillator. Earlier analyses considered the environment coupled to a small polaron as a traditional system of linear harmonic oscillators [9] though later improved approaches introduced nonlinear oscillators too [10]. Accordingly $p = \pm p_{gu}\sqrt{[1- (E_{gu}/4E_{JT})^2]}$ which obtains at $p_{gu}\times 2GQ/\sqrt{[(2GQ)^2 + E_{gu}^2]}$ for $Q = \pm Q_{min} = \pm\sqrt{[4G^4 - K^2 E_{gu}^2]}/2GK$ from the linear oscillator model [9], and $\mu = (4\pi\mu_0/c)(\pi\rho^2)\Omega_{rot}$ from the nonlinear model [11]. Here $p_{gu}$ is the electric dipole arising from the breakup of inversion symmetry at the small polaron site, Q is the coupled phonon mode coordinate, K is the stiffness, G is the electron-phonon coupling constant, $\mu_0$ is the magnetic permeability of the medium, $\rho$ and $\Omega_{rot}$ are the small polaron orbital radius and rotational frequency, respectively. The last two quantities are inherent to the nonlinear oscillator representing the small polaron. Under these conditions the electric dipole will be holding good at energies closer to the well bottom, while the magnetic dipole will hold true within most of the energy range.

Introducing the external field into the analysis extended so as to account for the field-coupling terms, we see that the main quantity to be affected by the applied field is the vibronic tunneling splitting $\Delta E$ which turns into

$$\Delta E(F) = \sqrt{[\Delta E(0)^2 + H'^2_F]} \tag{12}$$

where $\Delta E(0)$ is the field-off splitting by equation (3) and $H'_F$ is the field-coupling energy. At low fields $[H'_F \ll \Delta E(0)]$ $\Delta E(F) = \Delta E(0)\{1 + \frac{1}{2} [H'_F/\Delta E(0)]^2\}$, at high fields $[H'_F \gg \Delta E(0)]$ we have $\Delta E(F) = H'_F \{1 + \frac{1}{2} [\Delta E(0)/ H'_F]^2\}$. The field effect on the small-polaron superlattice converts the low-field condition to a high-field one. At any rate, the field-dependent tunneling splitting is superior to the field-off one: Consequently, from equation (6) the field-off pairing energy (6) is superior to the field-on one which implies that the field effect always tends to suppress the conductivity.

2.3. Polaron confinement models

2.3.1. Dipole-dipole coupling of hindered rotators

As stated before, the vibronic polaron size is determined by the ratio of two parameters: Jahn-Teller's energy $E_{JT}$ and the interlevel bandgap $E_{gu}$: namely $S = E_{gu} / 4E_{JT}$. Accordingly, vibronic polarons are *large* for $4E_{JT} \sim E_{gu}$ ($S \sim 1$), *intermediate* for $E_{gu} < 4E_{JT}$ ($S < 1$) and *small* for $E_{gu} \ll 4E_{JT}$ ($S \ll 1$). There is an additional requirement in that there should be at least one underbarrier level for the vibronic system to occupy: $E_B / \frac{1}{2}\hbar\omega = (2E_{JT}/\hbar\omega)[1-(E_{gu}/4E_{JT})] \geq 1$ in which $E_B$ is the barrier height. The adopted terminology is based on the relative magnitude of the electric conductivity, there being no distinction between electronic and large polaron conductivities, while the small polaron conductivity is virtually vanishing (localization). It follows that along the temperature axis the polarons are first large, then abruptly become small and confined following a structural transfiguring.

Nevertheless, the problem should be addressed on a statistical basis: starting with the free energy of a vibronic *polaron gas*. This free energy analysis has been done, as reported in Ref. [12]. The temperature dependence of the renormalized vibrational frequency $\omega_{ren}$ of a system of coupled vibronic rotators (polaron gas) has been derived, as reproduced in Figure 5. This is a series showing the softening of the frequency of an average hindered rotator as the temperature is raised with the pairing energy as parameter. It should be stressed that the polaron analysis leading to Figure 5 has accounted for elastic tunneling transitions only disregarding completely any possible inelastic contribution. The inelastic addenda to $\omega_{ren}$ are $\propto T^{1/2}$. If included, these would have undoubtedly improved the agreement with experiments within the thermally-activated range of Figure 5. Figure 5 shows the mode-softening temperature dependences at various pairing energies. Each one of these exhibits a constant mode frequency at the lowest temperatures followed by a decrement portion in which the frequency drops down possibly to values below those securing thermal occupation of an underbarrier level. If the remaining parameters are those of a small polaron (double well) potential, then we will have a small polaron materialized somewhere within the thermally activated range even it appears large before entering the thermally-activated range. Indeed, if the low temperature portion is populated by large polarons, the newly formed small polaron level will become populated as a result of the mode softening following the large to small polaron (low to high temperature species) conversion as in Figure 5. We see that, the Figure 5 result does not seem to easily fit the basic requirement if it accounts for soft modes appearing above 10 K in a low temperature large polaron moiety, the higher the pairing energy, the higher the mode softening temperature.

Alternatively, the mode softening points of the temperature dependences may be seen as indicative of higher temperature *polaron confinement*. Its origin is in the barrier controlled processes which become increasingly effective in confining the carriers as the temperature is raised. It is remarkable that the frequency within the lower temperature portion is constant which corresponds to a structureless large-polaron range. The small polarons fall into the thermally-activated range. This is a fundamental result of the confinement model: *a system of coupled hindered rotators with pairing interactions*. Presumably, the barrier height is ~ the pairing energy of the rotators. For an order of magnitude estimate, considering first-order pairing interactions alone may suffice.

2.3.2. Basic electrostatic equations

We consider it irrelevant whether the vibronic small polarons reside on a crystalline super-lattice or on some other type of periodic structure. The main requirement is that this structure be insulating and not conducting any electric currents. In this respect the prospects offered by the periodic solutions of the Poisson-Boltzmann (PB) equation are diverse. At the same time, Laplace's equations provide for solutions more remindful of periodic structures by alternating sign ions. In pursuing the optimal form, we shall make use of both equations.

2.3.2.1. Poisson-Boltzmann equation

The PB equation has been the subject of considerable attention for strong electrolytes and subsurface space charge effects in defect solids. Before going to the details, we introduce some relevant quantities [12]:

$\psi = e(\varphi - \varphi_\infty)/k_B T \rightarrow (e/kBT)(\varphi - \varphi_\infty)$ reduced electrostatic potential
$\varphi_\infty = (F^+ - F^-)/2e \rightarrow$ vanishing charge potential, $F^\pm$ formation free energies of $n^\pm$ particles
$\rho(\mathbf{r}) = -2e\, \tilde{N} \sinh(\psi) \rightarrow$ bulk (space) charge
$\tilde{N} = N \exp[-(F^+ - e\varphi_\infty)/k_B T] \rightarrow$ normalization factor, $N \rightarrow$ total number of sites
$\Delta\psi = \upsilon^2 \sinh\psi \rightarrow$ Poisson-Boltzmann equation  (13)

Last is the Poisson-Boltzmann equation obtained by inserting the bulk charge from equation (3) into Poisson's equation. $\upsilon = r_D^{-1} = \sqrt{(8\pi Ne^2/\kappa k_B T)}$, is the reciprocal Debye screening radius, $\kappa$ is the dielectric constant,. Here and above N stands for the total number of sites in the sea of fundamental particles. The quantity $\psi$ is the reduced potential relative to the 'vanishing charge potential' $\varphi_\infty$.

To solve for (5) in 1D we set $\psi(\mathbf{R}) = \psi_1(X)\psi_2(Y)\psi_3(Z)$ where $X = \upsilon x$, $Y = \upsilon y$, $Z = \upsilon z$; then

$$\Delta\psi = \sinh\psi \quad (14)$$

in capital (dimensionless) coordinates, $\Delta = \Sigma_i(\partial^2/\partial X_i^2)$.

For a system of plates (sheets), equation (14) reduce to the coupled equations to

$$\partial^2\psi/\partial Z^2 = \sinh\psi \tag{15}$$

$$\partial^2\psi/\partial X^2 + \partial^2\psi/\partial Y^2 = 0 \tag{16}$$

To solve for $\psi(Z)$ in (15) we multiply both sides by $d\psi/dZ$ and integrate using $(\partial/\partial Z)(\partial\psi/\partial Z)^2 \equiv 2(\partial\psi/\partial Z)(\partial^2\psi/\partial Z^2) = 2(\partial\psi/\partial Z)\sinh\psi$ wherefrom it follows [12]:

$$\partial\psi/\partial Z = \pm\sqrt{(2\cosh\psi + C)} = \pm 2\sqrt{[\cosh^2(\tfrac{1}{2}\psi) - k^2]} \tag{17}$$

where $C = 2(1-2k^2)$ is an integration constant, k is the elliptic integral modulus.

We solve for (15') at constant amplitudes to find that in 1D, the solutions for $\psi(Z)$ are either periodic or aperiodic functions, e.g.:

$$\psi_p(Z) = \ln\cotan(Z) \ (k = 0) \ \text{periodic solution} \tag{18}$$

$$\psi_p(Z) = \ln[cn(\tfrac{1}{2}Z,k)/sn(\tfrac{1}{2}Z,k)dn(\tfrac{1}{2}Z,k)] \ \text{general solution formula} \tag{19}$$

$$\psi_a(Z) = \ln\tanh(Z) \ (k = 1) \ \text{aperiodic solution} \tag{20}$$

It is easily seen that equations (15") are solved to give either screening tails or periodic solutions of the linear harmonic-oscillator type along either the X or Y axes. Both can be regarded as edge effects at the plate borders which plates arrange along the Z-axes in accordance with equations (15). See Figure 6 for graphs pertinent to the PB plates.

---

2.3.2.2. Laplace equations

In view of the point charge character of the host ions, 3D solutions to the Laplace equation (L) (vanishing space charge density) may be considered appropriate. However, due to the accompanying lattice distortions around the electronic charge the 3D Poisson-Boltzmann equation (finite bulk charge density) is more likely to reproduce the actual charge distribution at the polaron site. Unfortunately the latter (PB) are not available, while the former (L) gives

$$\Delta\psi = 0$$

$$\psi(\mathbf{r}) = \ln[(1+\lambda)/(1-\lambda)]^2 \tag{21}$$

---

$$\lambda = Au(\alpha x,k_1)v(\beta y,k_2)t(\gamma z,k_3) + B\underline{u}(\alpha x,k_1)\underline{v}(\beta y,k_2)\underline{t}(\gamma z,k_3) +$$

$$Cu^*(\alpha x,k_1)v^*(\beta y,k_2)t^*(\gamma z,k_3) \tag{22}$$

$u(\alpha x, k_1)$, etc. are some of 12 generated elliptic functions, $k_i$ are the elliptic integral moduli, A through C are constants [13]. Despite the lack of any 3D PB solution, the Laplace alternative may be plausible, especially when the small polaron condition is rigorously met. See Figure 7 for Laplace equation graphs.

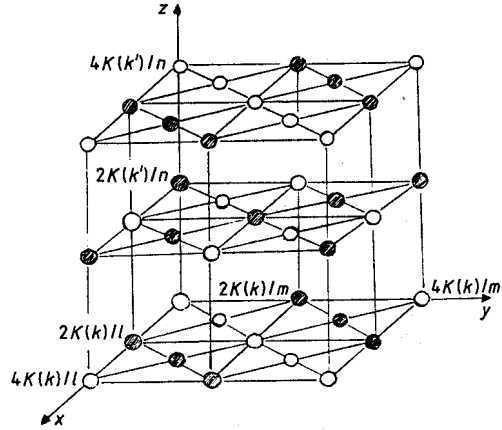

Figure 7: A 3D cubic lattice obtained from the analytic solution as dscribed by the first two terms in the solution (21) & (22) of Laplace's equation. (From Reference 23).

Generally, the solution to Laplace's equation is obtained through separating the variables. From (14)-(15) we have assuming $\psi(\mathbf{R}) = \psi_1(X)\psi_2(Y)\psi_3(Z)$

$$(1/\psi_1)(d^2\psi_1/dX^2) + (1/\psi_2)(d^2\psi_2/dY^2) = 0 \tag{23}$$

which is integrated to give the following independent though equivalent equations:

$$(1/\psi_1)(d^2\psi_1/dX^2) = [d(\ln\psi_1)/dX](d^2\psi_1/dX^2)/(d\psi_1/dX)$$

$$(1/\psi_2)(d^2\psi_2/dY^2) = [d(\ln\psi_2)/dY](d^2\psi_2/dY^2)/(d\psi_2/dY) \tag{24}$$

We set $\psi_1 = sn(X)$, $\psi_2 = cn(X)$ and insert into (13)

$dsn(X)/dX = cn(X)dn(X)$, $dcn(X)/dX = -sn(X)dn(X)$, $ddn(X)/dX = -k^2 sn(X)cn(X)$

to get:

$$(d^2sn(X)/dX^2)/sn(X) = -dn^2(X) - k^2 cn^2(X) \tag{25}$$

$$(d^2cn(Y)/dY^2)/cn(Y) = -dn^2(Y) - k^2 sn^2(Y) \tag{26}$$

The last two equations (22) and (23) are compatible at k = 0 which means trigonometry. Other choices are $\psi_1(X) = sn(X)$, $\psi_2(X) = sn(Y)$. They satisfy equation (13) outright.

Alternatively, equation (18) may be interpreted as follows:

$$(1/\psi_1)(d^2\psi_1/dX^2) = \pm C, (C > 0) \tag{27}$$

where C is an X & Y independent constant. We arrive at $\psi_1'' \pm C\psi_1 = 0$ which is readily solved to give $\psi_1(X) = \exp(\pm\alpha X)$ with $\alpha^2 = \pm C^{1/2}$. The solution again implies either a screening solution or a linear oscillator motion along each of the capital X,Y axes.
Turning back to the (21)&(22) forms, we checked the function

$$\psi = 4q\tanh^{-1}[cn(lx,k)cn(my,k)cn(nz,K) + 4q \tanh^{-1}[cn(lx,k)cn(my,k)cn(nz,k') +$$

$$sn((lx,k)sn(my,k)dn(nz,k')]$$

with periods $T_x = 4K(k)/l$, $T_y(4K(k)/m$, $T_z = 4K(k')/n$.

This 3D function is reproduced in Figure 7 for a Laplace structure.

2.3.2.3. Poisson-Fermi equations

Poisson-Fermi (PF) equations have appeared in relevance to fermion systems undergoing a quantum statistics. Quantum statistics conditions occur when the particle densities may no longer be considered small with respect to the site densities. PF may be regarded essential for solving statistics problems related to fermion systems. The PF equation reads [14,15]:

$$\Delta\psi = \sinh\psi / (1 + P^2 + 2P \cosh\psi) \tag{28}$$

where $P = \exp[-(F^+ - e\varphi_\infty)/k_BT]$. The Boltzmann tail "classic" statistics holds good for $P \ll 1$ ($F^+ \gg e\varphi_\infty$). No solutions to the PF equations have been reported so far. Nevertheless, for $e\varphi_\infty \sim F^+$ we have $P \sim 1$ and equation (23) reduces to

$$\Delta\psi \cong \sinh\psi / 2(1 + \cosh\psi) = \tfrac{1}{2}\tanh(\psi/2) \rightarrow \Delta(\tfrac{1}{2}\psi) \cong \tfrac{1}{4}\tanh(\tfrac{1}{2}\psi) \tag{29}$$

Equation (23) may be solved in 1D (plates) by a procedure similar to the one leading to equations (13): On multiplying by $d\psi/dZ$ and integrating we get

$$d\psi/dZ = \pm\sqrt{2} \int [\sinh\psi / (1 + P^2 + 2P\cosh\psi)]^{1/2} d\psi$$

$$d\psi/dZ = \pm 2^{-1/2} \int [\sinh\psi / (1 + \cosh\psi)]^{1/2} d\psi \quad (P = 1) \tag{30}$$

$$d\psi/dZ = \pm\sqrt{2} \int [\sinh\psi]^{1/2} d\psi \ (P = 0)$$

which may be integrated in elliptic functions. For $P \sim 1$ we get (2.464.5):

$$d(\tfrac{1}{2}\psi)/dZ \sim \pm \int\sqrt{[\sinh\psi /(1+\cosh\psi)]^{1/2}}d\psi = \pm \int \sqrt{[\tanh(\tfrac{1}{2}\psi)]}d(\tfrac{1}{2}\psi) \tag{31}$$

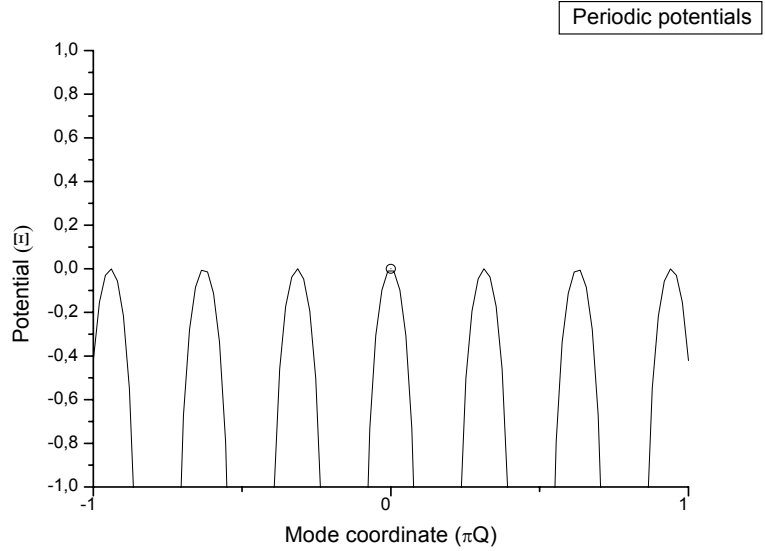

Figure 8: Periodic potentials ($\Xi = \sin(\psi)$) in the PF theoretical model. The present version provides an approximate solution to the Poisson-Fermi electrostatic equation holding better at low $\psi$.

The hyperbolic tangent is not tabulated in textbooks on integrals [16] which calls for approximations. One is $\tanh x \sim \sinh x$ which holds good at $\frac{1}{2}\psi \ll 1$ (P = 1):

$\pm \int \sqrt{[\tanh(\frac{1}{2}\psi)]}d(\frac{1}{2}\psi) \sim \pm \int \sqrt{[\sinh(\frac{1}{2}\psi)]}d(\frac{1}{2}\psi) =$

$\pm 4\{[F(\alpha,k) - 2E(\alpha,k)] + 4\sqrt{([\sinh(\frac{1}{2}\psi)][1+\sinh^2(\frac{1}{2}\psi)])} / [1+\sinh(\frac{1}{2}\psi)]\} + C$  (32)

with $\alpha = \cos^{-1}\{[1-\sinh(\frac{1}{2}\psi)] / [1+\sinh(\frac{1}{2}\psi)]\}$, $k = 1/\sqrt{2} = 0.7071$, $\frac{1}{2}\psi > 0$, $a = \frac{1}{4}$

while for P = 0 (2.464.5):

$d\psi/dZ = \pm\sqrt{2} \int \sqrt{[\sinh\psi]}d\psi =$
$\pm 4\{[F(\alpha,k) - 2E(\alpha,k)] + 4\sqrt{([\sinh(\psi)][1+\sinh^2(\psi)])} / [1+\sinh(\psi)]\} + C$

with $\alpha = \cos^{-1}\{[1-\sinh(\psi)] / [1+\sinh(\psi)]\}$, $k = 1/\sqrt{2} = 0.7071$, $\psi > 0$, $a = \frac{1}{2}$  (33)

Examples for fermion systems in solids are provided by conduction electrons in metals below the Fermi energy (temperature). For lower conversion temperatures below, say 100 K, the PF equation may provide a better approach to reality, as may fermion systems. If so, the periodic solutions for fermions may be found more reliable.

Now, we obtain for P = 1:

---

$$Z = \int d\psi / \pm 4\{[F(\alpha,k) - 2E(\alpha,k)] + 4\sqrt{[\sinh\tfrac{1}{2}\psi]} \times (\cosh\tfrac{1}{2}\psi) / (1+\sinh\tfrac{1}{2}\psi) + C \quad (P=1)(34)$$

which is to be solved for the purpose. For P = 0 we get likewise

$$Z = \int d\psi / \pm 2\sqrt{2}\{[F(\alpha,k) - 2E(\alpha,k)] + 2\sqrt{[\sinh\psi]} \times (\cosh\psi) / (1+\sinh\psi)\} + C \quad (P=0) \ (35)$$

Next we perform substitutions leading to equations for Z:

$$\alpha = \cos^{-1}\{[1-\sinh(\tfrac{1}{2}\psi)] / [1+\sinh(\tfrac{1}{2}\psi)]\}, \ k = 1/\sqrt{2} = 0.7071 \quad (P=1) \qquad (36)$$

$$\alpha = \cos^{-1}\{[1-\sinh(\psi)] / [1+\sinh(\psi)]\}, \ k = 1/\sqrt{2} = 0.7071 \quad (P=0) \qquad (37)$$

Taking the cosine on both sides we get

$$\cos^2\alpha = [(1-\sinh\tfrac{1}{2}\psi)/(1+\sinh\tfrac{1}{2}\psi)]^2, \ \sin^2\alpha = 1-[(1-\sinh\tfrac{1}{2}\psi)/(1+\sinh\tfrac{1}{2}\psi)]^2 \ (P=1) \ (38)$$
$$\cos^2\alpha = [(1-\sinh\psi)/(1+\sinh\psi)]^2, \ \sin^2\alpha = 1-[(1-\sinh\psi)/(1+\sinh\psi)]^2 \quad (P=0) \ (39)$$

$$\cos^2\alpha = [(1-\sinh\tfrac{1}{2}\psi)/(1+\sinh\tfrac{1}{2}\psi)]^2 \ (P=1)$$
$$\cos^2\alpha = [(1-\sinh\psi)/(1+\sinh\psi)]^2 \ (P=0) \qquad (40)$$

$$1-2[\cos\alpha+1]^{-1} = \sinh(\tfrac{1}{2}\psi), \ \psi = 2\sin^{-1}(1-2[\cos\alpha+1]^{-1}) \ (P=1)$$
$$1-2[\cos\alpha+1]^{-1} = \sinh(\psi), \ \psi = \sin^{-1}(1-2[\cos\alpha+1]^{-1}) \ (P=0) \qquad (41)$$

Equations (37)&(38) give the relationship between $\psi$ and $\alpha$ for P=1 and P=0 upper and lower; respectively, so do (27) and (28) between Z and $\psi$ for P=1 and P=0, respectively. Making use of the definitions, we obtain the basic elliptic functions: $cnZ = \cos\alpha$, $snZ = \sin\alpha$, $dnZ = \sqrt{(1-k^2\sin^2\alpha)}$ and the basic elliptic integrals: $F(\alpha,k) = \int_0^\alpha d\varphi / \sqrt{(1-k^2\sin^2\varphi)} \to 1^{st}$ kind, as well as $E(\alpha,k) = \int_0^\alpha d\varphi \sqrt{(1-k^2\sin^2\varphi)} \to 2^{nd}$ kind and combine them to construct solutions $\psi = \psi(Z)$ to the PF equations. See Figure 8 for PF graphic illustrations.

Manipulating further we get from (38)

$$Z = \int d\psi / \pm 4\{[F(\alpha,k) - 2E(\alpha,k)] + 4\sqrt{[\sinh\tfrac{1}{2}\psi]} \times (\cosh\tfrac{1}{2}\psi) / (1+\sinh\tfrac{1}{2}\psi)\} + C \ (P=1) \ (42)$$
$$\psi = 2\sin^{-1}(1-2[\cos\alpha+1]^{-1}) \text{ or } \cos\alpha = \pm[(1-\sinh\tfrac{1}{2}\psi)/(1+\sinh\tfrac{1}{2}\psi)] \ (P=1)$$

From the above we obtain the solutions by periodic functions cn(Z,k):

$$\psi(Z,k) = 2\sin^{-1}(1-2[cn(Z,k)+1]^{-1}) \ (P=1). \qquad (43)$$

Further, we get likewise for P = 0:

$$Z = \int d\psi / \pm 2\sqrt{2}\{[F(\alpha,k) - 2E(\alpha,k)] + 2\sqrt{[\sinh\psi]} \times (\cosh\psi) / (1+\sinh\psi) + C \ (P=0) \ (44)$$
$$\psi = \sin^{-1}(1-2[\cos\alpha+1]^{-1}) \text{ or } \cos\alpha = \pm[(1-\sinh\psi)/(1+\sinh\psi)] \ (P=0)$$

---

while the periodic solution is

$$\psi(Z,k) = \sin^{-1}(1-2[cn(Z,k) + 1]^{-1}) \quad (P=0). \tag{45}$$

Fimally, we also rearrange $E(\alpha,k)$ to get

$$E(\alpha,k) = {}_0\!\int^{\alpha} d\varphi\, (1 - k^2\sin^2\varphi)/\sqrt{(1 - k^2\sin^2\varphi)} =$$

$$F(\varphi,k) - \int d\varphi\, k^2\sin^2\varphi/\sqrt{(1 - k^2\sin^2\varphi)} = F(\varphi,k) - F(\varphi,k^2) - E(\varphi,k^2) \quad (2.584.1,4) \tag{46}$$

### 2.3.4. Electrostatic strings (PB)

In addition to periodic structures of crystalline character, the small polaron may form string-like periodic structures in electrostatic systems. The strings being 1D units, they reduce to 2D the requirement for a maximum dimension of the PB equation. These strings extend along the z-axis and are each equipotential. An example is the nonlinear periodic function [17,18]:

$$\psi_{strings} = 4\tanh^{-1}[cn(\alpha x, k_1)cn(\alpha y, k_2)], \qquad k_1^2 + k_2^2 = 1 \tag{47}$$

which has solved the 2D PB equation. See Figure 9 for a graph pertinent to that equation.

### 2.3.5. Electrostatic plates (PB)

Electrostatic units of planar symmetry are obtainable too. If these are planes, they reduce the maximum dimension to 1D. The 2D planes are equipotential plates of size exceeding largely the Debye screening length. Check Figure 6 for examples of a series of equipotential plates (sheets) with an increasing development along the z-axis which solve the 1D PB equation:

$$\psi_{plates}(Z) = \ln[cn(\tfrac{1}{2}Z,k)/sn(\tfrac{1}{2}Z,k)dn(\tfrac{1}{2}Z,k)] \quad \text{general plate formula.} \tag{48}$$

### 2.3.6. Free energy analysis

The free energy of a small polaron moiety is $\Im = U - TS$ with internal energy

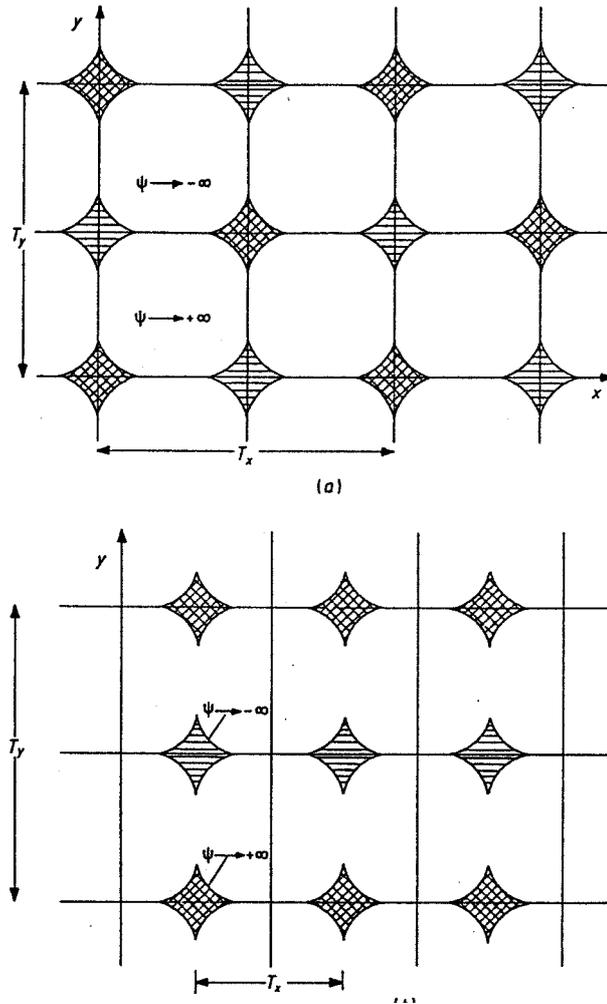

Figure 9: 2D periodic arrangements described by the electrostatic equations (PB and L alike). These pertain to the (X,Y) surface, but if they do, they can be extended in-depth in the form of equipotential strings along the Z-axis. In as much as neighboring ionic islands display alternate positive and negative electric charges, the resulting 3D structure corresponds nearly perfectly to what can be expected for an array of electrostatic strings.(From Reference 18).

$$U = \int d\mathbf{r}\, [F^+ n^+ + F^- n^-] \qquad (49)$$

and configurational entropy at $n/N \ll 1$:

$$S = -k_B \int dr \{n^+[\ln(n^+/N) - 1] + n^-[\ln(n^-/N) - 1]\} \qquad (50)$$

where the particle densities are assumed to follow the Boltzmann-tail statistics in an external field $\mathbf{F} = -\nabla\varphi(\mathbf{r})$:

$$n_{\pm}(\mathbf{r}) \sim n_0 \exp(\pm e\varphi(\mathbf{r})/k_B T) \qquad (51)$$

with $n_0 = \exp[-\mu/(k_B T)]\}$ ($\mu$ is the chemical potential or Fermi's energy) to determine the bulk charge. Inserting we get

$$\Im(Q) = -k_B T \ln\{2[1+\cosh([(\tfrac{1}{2}E_{gap})^2 + (GQ)^2]^{1/2}/k_B T)]\} + \tfrac{1}{2}KQ^2 + U(R,Q) \qquad (52)$$

where Q is the radial phonon coordinate, $K = M\omega_{bare}^2$, $U(R,Q)$ is the pair interaction function at R. For a weak dipole-dipole coupling of vibronic oscillators,

$$U(R,Q) = u(R)v(Q) = [R^2(\mathbf{p}_1.\mathbf{p}_2)-3(\mathbf{p}_1.\mathbf{R})(\mathbf{p}_2.\mathbf{R})]/\kappa R^5 \times$$

$$4G_1Q_1G_2Q_2/\{[(2G_1Q_1)^2+(\tfrac{1}{2}E_{gap})^2][(2G_2Q_2)^2+(\tfrac{1}{2}E_{gap})^2]\} \qquad (53)$$

$Q_i$ and $G_i$ are the coordinate and mixing constant for the i-th mode.

The derivatives of the free energy might also be useful for further analyses:

$$\Im'(Q) = KQ - \tanh(\sigma/2)(G^2/k_B T)(Q/\sigma) + 2u(R)G^2(\tfrac{1}{2}E_{gap})^2 Q/(\sigma k_B T)^4$$

$$\Im''(Q) = K\{1 - G^2/Kk_B T[\sigma^{-1}\tanh(\sigma/2) +$$

$$(GQ)/(\sigma k_B T)^2[1/2(\cosh(\sigma/2))^2 -$$

$$(1/\sigma)\tanh(\sigma/2)]]\} - 2u(R)G^2(\tfrac{1}{2}E_{gap})^2[3(GQ)^4 + 2(GQ)^2(\tfrac{1}{2}E_{gap})^2 -$$

$$(\tfrac{1}{2}E_{gap})^4]/(\sigma k_B T)^8 \qquad (54)$$

where we have introduced the notation $\sigma = [\tfrac{1}{2}(E_{gap})^2 + (GQ)^2]^{1/2}/k_B T$ for brevity.

To derive a renormalized frequency, we also note that $1/2E_{JT} = \tanh(\sigma/2)/(\sigma k_B T) - 2u(R) \times (\tfrac{1}{2}E_{gap})^2/(\sigma k_B T)^4$ holds good for the side minima at $Q = Q_0$ where

$$Q_0 = \{(2E_{JT}/K)[1 - (E_{gap}/4E_{JT})^2]\}^{1/2} \qquad (55)$$

We then obtain

$$\Im''(Q_0) = K\{1 - (2E_{JT}/k_B T)[(1/\sigma)\tanh(\sigma/2)+(GQ_0)^2/(\sigma k_B T)^2 \times$$

$$[1/2[\cosh(\sigma/2)]^2 - (1/\sigma)\tanh(\sigma/2)]] - 4u(R)E_{JT}(\tfrac{1}{2}E_{gap})^2 \times$$

$$[3GQ)^4 + 2(GQ)^2(\tfrac{1}{2}E_{gap})^2 - (\tfrac{1}{2}E_{gap})^4]/(\sigma k_B T)^8\} \qquad (56)$$

wherefrom the renormalized frequency follows suit

$$\omega_{ren} = \omega_{bare}\sqrt{[\Im''(Q_0)/K]} \qquad (57)$$

It may also be instructive finding the temperature $T_C$ of high-symmetry to low-symmetry conversion from $\Im''(0) = 0$:

$$\Im''(0) = K\{1 - (2G^2/KE_{gap})[\tanh(E_{gap}/4k_BT) - 2u(R)/(\tfrac{1}{2}E_{gap})]\} \qquad (58)$$

The conversion temperature $T_C$ reads

$$k_BT_C = (\tfrac{1}{4}E_{gap})/\tanh^{-1}\{(E_{gap}/4E_{JT})[1 + 2u(R)/(\tfrac{1}{2}E_{gap})\times(4E_{JT}/E_{gap})]\} \qquad (59)$$

We next compare the free energies of each of the strings and plates with the free energy of the random charge distribution, as shown elsewhere for the plates. We make this in order to see which one (strings, plates) are more favorable from the statistical point of view to confining the small polarons. The result can be seen in Figure 10.

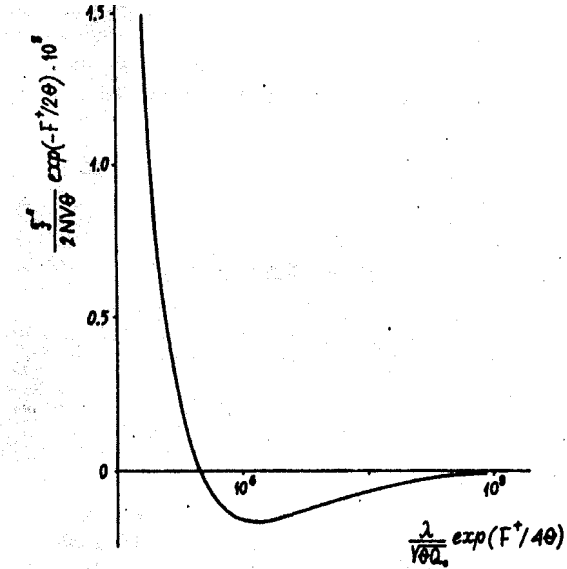

Figure 10: Free energy of a 1D planar PB structure (2D plate) in appropriate coordinates.

2.3.7. Electrostatic energies

The above equation (17) relates to the free energy of a random particle distribution (RPD) in so far as it does not include any electrostatic potential containing term. Such terms will arise from entering the contribution of the subsurface space charge and/or the periodic part of the solution. Once these contributions are available. the free energies of pairs, plates and strings may be evaluated and compared with each other. The electrostatic contribution will be evaluated as energy of the field using:

$$V(Z) \sim \int(\varepsilon/2)[\mathbf{F(Q,Z)}]^2 \, d\mathbf{r} = \int(\varepsilon/2)[-\nabla\varphi\mathbf{(Q,Z)}]^2 \, d\mathbf{r} \tag{60}$$

2.3.7.1. Plates

Inserting the least period periodic solution from (8') into equation (23) we obtain

$$V(Z) = \int(\varepsilon/2)[-d\varphi(z)/dz]^2 dz = \upsilon^{-4}\int(\varepsilon/2)(k_BT/e)^2[-d\psi(Z)/dZ]^2 dZ$$

$$= \varepsilon(k_BT/e)^2\upsilon^{-4}\int \sin(Z)dZ = 2\varepsilon(k_BT/e)^2\upsilon^{-4} \quad (24) \tag{61}$$

It is not surprising to get free energies of the order of $k_BT$, in so far as the expectancy is of a low energy field contribution.

2.3.7.2. Strings

We take as an example for strings the 2D nonlinear periodic function from equation (13):

$$\psi = 4\tanh^{-1}[cn(\alpha x,k_1)cn(\alpha y,k_2)], \qquad k_1^2 + k_2^2 = 1 \tag{62}$$

$$\psi'_u = 4(1-u^2) = \tfrac{1}{2}\ln[(1+u)/(1-u)], \quad u = cn(\alpha x,k_1)cn(\alpha y,k_2) \tag{63}$$

---

to obtain

$$V(x,y) = (\varepsilon k_BT/2e)\int[(\partial u/\partial x)^2 + (\partial u/\partial y)^2]\{\tfrac{1}{2}\ln[(1+u)/(1-u)]\}^2 dxdy \tag{64}$$

with

$$\partial u/\partial x = \{[\partial cn(\alpha x,k_1)/\partial x]cn(\alpha y,k_2) + [\partial cn(\alpha x,k_2)/\partial x]cn(\alpha y,k_1)\}$$

$$\partial u/\partial y = \{[\partial cn(\alpha y,k_1)/\partial y]cn(\alpha y,k_2) + [\partial cn(\alpha y,k_2)/\partial y]cn(\alpha x,k_1)\}. \tag{65}$$

2.3.8. Tunneling transitions across barriers in electrostatic plates

There is only one work available so far to put up the question of tunneling transitions and associated energy bands in periodic PB or PF structures [12,17]. This work most certainly contributes to the physics and mathematics of periodic plates and sets up a precedent for studying other kinds of periodic structures. Among other things, it shows that the energy of a fermion particle intruding from outside to the confinement region is quantized into allowed energy bands. It is very likely that there will be a barrier preventing that particle from leaving the periodic structure (confinement).

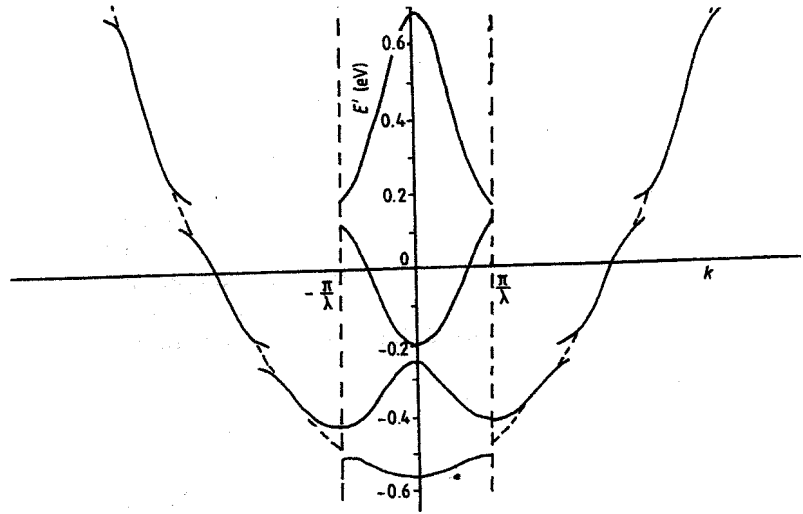

Figure 11: Energy bands structure predicted by the PB equation. The following parameters have been used: $e\varphi_\infty = 054$ eV, $F+ = 0.7$ eV, $k_BT = 0.025$ eV, $\varepsilon = 4.8$, $r_D = 6.28$Å, $m = 0.6m_e$. (From Reference 17).

3. Discussion

We consider the present Poisson-Fermi equation as a sole result obtained by combining the classical Poisson equation with a quantum statistical particle distribution. Earlier discussed equations have emerged as products of combining Poisson's equation with classical particle distributions, Boltzmann tail ones, to be exact. Interpreting the classical statistics on a quantum basis opens the way for comparing classical and quantum results on a common platform.

The basic electrostatic statistical equations (PB and PF), as well as the basic electrostatic equation (L) have already found a wider use for promoting a better understanding of the distribution of point defects in crystals, the occurrence of subsurface space charges, and now even the formation of strings which will seemingly play an increasing role in solid state and nuclear physics. For this reason PB's and PF's periodic solutions will still be sought here and there in combination with classical statistics (Boltzmann tail) and quantum statistics (Fermi – Dirac). We are reserved towards Bose-Einstein statistics compact particles insofar as there may not be any genuine bosons but rather fermion pairs, Cooper's or real space ones. Some implications for the electrostatic equations in higher symmetry are discussed elsewhere [18-21]. There are a few other confinement events considered so far in the literature. Such is the color confinement which immobilizes color quarks during the initial stages of particle generation processes following the big bang. In as much as fermions and bosons alike may not be active in the absence of quarks, the confinement is particle active too.

## 4. Conclusion

We have investigated various proposals to effect small polaron confinement in solids, such as forming 3D-crystal like structures as well as 2D-electrostatic strings and 1D-electrostatic plates. The discrete crystallographic structures are described by periodic solutions to the Laplace equation, the quasi continuous strings are by periodic solutions to the Poisson-Boltzmann (PB) equation. Another possibility has also been investigated is the Poisson-Fermi equation (PF) which is likely to produce a better string approach to fermion systems, though most of the related work will be left for a subsequent study.

Acknowledgement. This paper is dedicated to the memory of my longtime friend and co-author D. Ouroushev. Without his devotion, the field of electrostatic equations would have never reached maturity.

References

[1] A.J. Millis, Nature **392**, 147 (1998).
[2] P.W. Anderson, "*The Theory of The Superconductivity in The High-Tc Cuprates*" (Princeton Studies in Physics, Princeton NJ, 1998).
[3] J. Georg Bednorz and K. Alex Mueller,. (1988). Nobel Lecture in Physics, Revs. Modern Phys. **60**, 585-600 (1987):
   "*Perovskite-type oxides - the new approach to high-Tc superconductivity.*"
[4] J. Fontcuberta, Physics World, February, 33-38 (1999).
[5] C. Joos, L. Wu, T. Beetz, R.F. Klie, M. Beleggia, M.A. Schofield, S. Schramm, J. Hoffmann, and Y. Zhu, PNAS 104 (34) 13587-13602 (2007).
[6] M. Georgiev and M. Borissov, Phys. Rev. B **39** (16) 11624-11632 (1989).
[7] M. Georgiev and M. Borissov, cond-mat/0601423.
[8] I.B. Bersuker, *The Jahn-Teller Effect and Vibronic Interaction in Modern Chemistry* (Academic, New York, 1984).
[9] S.G. Christov, Phys. Rev. B **26** (12) 6918-6935 (1982).
[10] G. Baldacchini, R.M. Montereali, U.M. Grassano, A. Scacco, P. Petrova, M. Mladenova, M. Ivanovich, and M. Georgiev, cond-mat 0709.1951, physics 0710.0650, physics 0710.2334, physics 0710.3583; F. Bridges, CRC Critical Revs. **5**, 1-88 (1975).
[11] A.G. Andreev, M. Georgiev, M.S. Mladenova, V. Krastev, Internat. J. Quantum Chem. **89**, 371-376 (2002). Petrova *et al.* in: *Quantum Systems in Chemistry and Physics* (Kluwer, Utreht (1997) 373-395.

[12] M. Georgiev, N. Martinov, and D. Ouroushev, Crystal Latt. Def. **9**. 1-11 (1980).
[13] N. Martinov and D. Ouroushev, Materials Science **16** (4) 47-51 (1990).
[14] M. Georgiev, N. Martinov, D. Ouroushev, Annuire de L'Universite de Sofia "St. Kliment Ohridski" **81**, 83-100 (1993).
[15] M. Georgiev, N. Martinov, D. Ouroushev, Annuire de L'Universite de Sofia "St. Kliment Ohridski" **81**, 101-110 (1993).
[16] Gradsteyn/Ryzhik Tables of Integrals, Series, and Products 5$^{th}$ edition (Alan Jeffrey ed.) (Academic Press, San Diego, 1994).
[17] N. Martinov, D. Ouroushev and M. Georgiev, J. Phys. C: Solid State Phys. **17**, 5175-5184 (1984).
[18] N. Martinov and D. Ouroushev, J. Phys. A: Math.Gen. **19**, 2707-2714 (1986).
[19] A.J.M. Garrett and L. Poladian, Annals of Phys. **188** (2) 386-434 (1988).
[20] N. Martinov and N. Vitanov, J. Phys. A: Math.Gen. **25**, L51-L56 (1992).
[21] N. Martinov, D. Ouroushev and A. Grigorov, J. Phys. A: Math.Gen. **24**, L975 (1991)
[22] M. Georgiev, physics arXiv:0711.4601; cond-mat 0712.3803.
[23] D. Ouroushev, J. Phys. A: Math. Gen. **18**, L845-L848 (1985).